\documentclass[conference]{IEEEtran}
\IEEEoverridecommandlockouts

\usepackage{cite}
\usepackage{amsmath,amssymb,amsfonts}
\usepackage{algorithmic}
\usepackage{graphicx}
\usepackage{textcomp}
\usepackage{todonotes}
\usepackage{graphicx}
\usepackage{lscape}
\usepackage{upgreek}
\usepackage{caption} 
\usepackage{hyperref}
\usepackage{todonotes}
\usepackage{amssymb}
\usepackage{lineno}
\usepackage{url}
\usepackage{array}
\usepackage{enumitem}
\usepackage{multirow}
\usepackage{soul}
\usepackage{float}
\usepackage{color}
\usepackage{pgfplots}
\usepackage{booktabs}
\usepackage[export]{adjustbox}
\usepackage{graphicx}
\usepackage{colortbl}
\usepackage{array}
\usepackage{fontawesome}
\usepackage{booktabs}
\usepackage[normalem]{ulem}

\usepackage{xcolor}
\usepackage[many]{tcolorbox}

\def\BibTeX{{\rm B\kern-.05em{\sc i\kern-.025em b}\kern-.08em
    T\kern-.1667em\lower.7ex\hbox{E}\kern-.125emX}}

\newtcolorbox{boxC}{
    colback = sub, 
    boxrule = 0pt,  
}

\definecolor{main}{HTML}{CFCFCF}    
\definecolor{sub}{HTML}{CFCFCF}     

\definecolor{darkmagenta}{rgb}{0.55, 0.0, 0.55}
\definecolor{darkgreen}{RGB}{6, 46, 3}
\definecolor{amber}{rgb}{1.0, 0.75, 0.0}
\definecolor{ao(english)}{rgb}{0.0, 0.5, 0.0}

\newcommand{\camera}[1]{{\color{black}#1}}

\usepackage{color}

\begin{document}

\title{Toward Organizational Decoupling in Microservices Through Key Developer Allocation}

\author{Xiaozhou Li$^1$, Noman Ahmad$^1$, Tomas Cerny$^2$,   Andrea Janes$^3$, Valentina Lenarduzzi$^1$, Davide Taibi$^1$ \\
$^1$\textit{University of Oulu}  --- 
$^2$\textit{University of Arizona}  --- 
$^3$\textit{Free University of Bozen-Bolzano}
\\
xiazhou.li@oulu.fi; noman.ahmad@oulu.fi
tcerny@arizona.edu \\ andrea.janes@unibz.it; valentina.lenarduzzi@oulu.fi; davide.taibi@oulu.fi
}


\maketitle

\begin{abstract}

With microservices continuously being popular in the software architecture domain, more practitioners and researchers have begun to pay attention to the degradation issue that diminishes its sustainability. One of the key factors that causes the degradation of software architecture is its organizational structure, according to Conway's Law. However, the best practice of ``One microservice per Team'', advocated widely by the industry, is not commonly adopted, especially when many developers contribute heavily across multiple microservices and create organizational coupling. Therein, many key developers, who are responsible for the majority of the project work and irreplaceable to the team, can also create the most coupling and be the primary cause of microservice degradation. Hence, to properly maintain microservice architecture in terms of its organizational structure, we shall identify these key developers and understand their connections to the organizational coupling within the project. We propose an approach to identify the key developers in microservice projects and investigate their connection to organizational coupling. The approach shall facilitate the maintenance and optimization of microservice projects against degradation by detecting and mitigating organizational coupling.

\end{abstract}


\section{Introduction}
\label{sec:Introduction}
As microservices continue to be the dominant architectural paradigm in software development, both practitioners and researchers have increasingly focused on the issue of architectural degradation, which undermines the long-term sustainability of such systems. A significant contributor to architectural degradation is the misalignment between the software's architectural structure and its organizational structure, as described by Conway's law. Despite the industry-wide promotion of the best practice of ``One microservice per team,'' it is often not implemented, particularly in scenarios where numerous developers contribute extensively to multiple microservices, resulting in organizational coupling. In such cases, key developers are then the ones who handle a significant portion of the project's work and are often indispensable to the team. Therefore, they can inadvertently introduce substantial coupling, becoming a primary cause of microservice degradation. As a consequence, to effectively preserve the sustainability of microservice architectures, it is essential to identify these key developers and analyze their contributions to organizational coupling within the project.

Therefore, we conceptualize an approach for the continuous maintenance and optimization of the organizational structure in microservice-based projects. The approach focuses on the identification of key developers and an analysis of their behavior to assess their influence on organizational coupling. Organizational coupling (OC) in microservice-based software projects is often defined as the extent to which two distinct microservices share developers who contribute significantly and alternately to both \cite{li2023evaluating}. By minimizing organizational coupling and refining the organizational structure, this work seeks to mitigate the degradation of microservice architecture from an organizational standpoint.

In this paper, we propose automated approaches to identify the different key developers in microservice-based projects from the project's commit and issue activity data. The early results demonstrate the usefulness of the approaches and the potential connection between the key developers and the organizational coupling issues.



\textcolor{black}{\textbf{Paper structure}. Section~\ref{sec:Design} introduces the study design, including the goals and research questions, as well as the relevant concepts and details in data collection and analysis methods. Moreover, Section~\ref{sec:Results} provides the early result, which is further discussed in Section~\ref{sec:Discussion}. Section~\ref{sec:RW} introduces the related work when Section~\ref{sec:Conclusion} concludes the paper.}

\section{Study Design}
\label{sec:Design}

In this section, we describe the study design, report the goal, research questions, data collection, and data analysis. To allow replicability and verifiability, we made public the raw data in the replication package hosted on Zenodo\footnote{\url{https://doi.org/10.5281/zenodo.14523750}}. 

\subsection{Goal and Research Questions}

The goal of this paper is to propose an approach to continuously maintain and optimize the organizational structure of microservice-based projects by identifying the various key developers and evaluating their behaviors toward causing organizational coupling. Furthermore, by maintaining the organizational structure via reducing organizational coupling, we shall contribute to mitigating the degradation of microservice architecture from the organizational perspective. 

Then, we first formulate an initial method for identifying the key developers in microservice projects. We adapt the method proposed by \cite{ccetin2020identifying,ccetin2022analyzing} into microservice projects when both the local key developers for each microservice and the global key developers across the whole project can be identified. 

Furthermore, by adopting the evaluation method of organizational coupling \cite{li2023evaluating}, we can further investigate the connection between identified key developers and their contribution to organizational coupling. Such a result shall lay the ground for future studies on constructing the corresponding decoupling strategies for different types of key developers.

With the knowledge obtained in the previous steps, practitioners, especially critical decision-makers, can suggest task allocation to the developers targeting organizational structural optimization. In addition, they shall also be supported by other context knowledge obtained from personality analysis, social network analysis, temporal network analysis, and so on \cite{li2023analyzing,li2024toward,bakhtin2024temporal}. This paper joins the other studies as one of the critical steps toward a concrete framework of microservice organizational structure optimization \cite{li2024framework}.

Therefore, we define two research questions (RQs): 

\begin{boxC}
\textbf{RQ$_1$}.How to identify the different key developers in microservice-based projects?
\end{boxC}

\begin{boxC}
\textbf{RQ$_2$}.How to assess key developers's contribution to organizational coupling?
\end{boxC}

\subsection{Key Developer Identification}

To identify the key developers in microservice-based projects, we adopted the approaches and concepts proposed by \c{C}etin and T\"uz\"un \cite{ccetin2020identifying,ccetin2022analyzing}. The authors propose to use \textit{artifact traceability graphs} to represent the interconnections between developers and the artifacts in any repository, e.g., commits and issues when all edges amongst artifacts and developers are undirected. Therein, the \textit{distance} between any artifacts (length of the edges) is calculated as 

\begin{equation}
    distance = \frac{1}{1 - \frac{\#days\: passed }{\#days\: in\: graph}}
\end{equation}


By constructing the artifact traceability graph of any given repository, three types of key developers can be identified respectively:

\begin{itemize}
    \item \textit{Jack} - a developer who has a broad knowledge of the project.
    \item \textit{Maven} - a developer who is a master in details of specific modules or files in the project.
    \item \textit{Connector} - a developer who is involved in different (sub)projects or different groups of developers.
\end{itemize}

To identify \textit{Jack} developers in the project, we need to calculate the file coverage ratio of each developer therein. Any file in the project is defined as \textit{reachable} by a particular developer when the artifact related to it is connected to the developer within the constructed graph under a certain distance threshold (the default threshold is set as $5$ by \cite{ccetin2020identifying}). However, a file is unreachable when the path contains another developer \cite{ccetin2020identifying}. Hence, the file coverage of any developer is simply the ratio of the number of reachable files to the number of all files in the project. Therefore, the \textit{Jack} developers are identified as the ones who have high file coverage.

To identify \textit{Maven} developers, we need to first find the \textit{rarely reached files} in the project, which means the files only reached by a limited number
of developers (set as reached by only $1$ developer in \cite{ccetin2020identifying}). Therefore, for each developer, his/her mavenness value is the ratio of the number of rarely reached files by him/her to the number of all rarely reached files when, hence, the {\textit{Maven}  developers can be identified as the ones who have high mavenness value.

To identify \textit{Connector} developer, we use the \textit{betweenness centrality} metric as the identifier \cite{freeman2002centrality}. This metric is commonly used to identify critical nodes in social networks and is also adopted to analyze developers' characteristics and behaviors in microservice projects \cite{li2023analyzing,li2024toward}.

\subsection{Organizational Coupling Evaluation}

Organizational coupling (OC), in terms of microservice-based software projects, is commonly seen as the degree to which two different microservices have common developers contributing to both heavily and alternately \cite{li2023evaluating}. 
One critical concept influencing the OC assessment is the developers' \textit{contribution switch} between microservices. This implies that the more frequently developers switch between microservices, the more heavily these two microservices are organizationally coupled. Therefore, by using the contribution switch as the weight, we can calculate the organizational coupling between two microservices as the sum of all common developers' average contributions between them, multiplied by the contribution weight \cite{li2023evaluating}.

Specifically, given two microservices $M_a$ and $M_b$ of a project, we assume a developer $D$ commits to both microservices. A contribution switch occurs when $D$ commits to $M_a$ and then commits to $M_b$ afterward or vice versa. Regarding the situation of logically coupled commits \cite{d2023microservice} where both microservices are changed in a single commit, two contribution switches are counted for this situation. Hence, the contribution switch for $D$ between $M_a$ and $M_b$, denoted as $S_{D}(M_{a}, M_{b})$ is calculated as follows:

\begin{equation}
    S_{D}(M_{a}, M_{b}) = \frac{k}{2\times{(n-1)}}
\end{equation}

\noindent where $k$ denotes the total number of contribution switches and $n$, is the total number of commits made by $D$ to both microservices.

Furthermore, \(OC(D, M_{a}, M_{b})\) defines the organizational coupling caused by $D$'s contributions across the two microservices. Meanwhile, \(\{ca_1, ca_2, ... ca_m\}\) represents the sequence of contribution values (e.g., LOC) 
for \(m\) commits made by $D$ to microservice \(M_a\) when \(\{cb_1, cb_2, ... cb_n\}\) represents $n$ commits by $D$ to \(M_b\). Hence, \(OC(D, M_{a}, M_{b})\) is calculated as follows,

\begin{equation}
    OC(D, M_{a}, M_{b}) = (\frac{2\sum_{j=1}^{m}ca_j\sum_{k=1}^{n}cb_k}{\sum_{j=1}^{m}ca_j + \sum_{k=1}^{n}cb_k}) \times S_{D}(M_{a}, M_{b})
\end{equation}

Therefore, we can calculate the total OC contribution of any $D$ by adding up his/her OC contribution in each microservice pair where he/she switches between.

\begin{table*}[!ht]
    \centering
    \adjustbox{max width=\textwidth}{
    \begin{tabular}{l|l|l|l}
    \hline
       \textbf{Service} & \textbf{Jack} & \textbf{Maven} & \textbf{Connector} \\
    \hline
    audit	&	[('sunXXX', 0.53), ('surXXX', 0.47), ('yesXXX', 0.13)]	&	[('sahXXX', 0.54), ('sunXXX', 0.46)]	&	[('sunXXX', 0.53), ('surXXX', 0.47), ('yesXXX', 0.13)] \\
authentication	&	[('yesXXX', 0.25), ('arpXXX', 0.21), ('sahXXX', 0.17)]	&	[('yesXXX', 0.3), ('sahXXX', 0.25), ('arpXXX', 0.2)]	&	[('yesXXX', 0.25), ('arpXXX', 0.21), ('sahXXX', 0.17)] \\
bpmn	&	[('surXXX', 0.11), ('yesXXX', 0.08), ('sahXXX', 0.08)]	&	[('surXXX', 0.47), ('sahXXX', 0.27), ('arpXXX', 0.13)]	&	[('surXXX', 0.11), ('yesXXX', 0.08), ('sahXXX', 0.08)] \\
chat	&	[('sahXXX', 0.11), ('yesXXX', 0.06), ('arpXXX', 0.04)]	&	[('sahXXX', 0.64), ('sunXXX', 0.18), ('yesXXX', 0.09)]	&	[('sahXXX', 0.11), ('yesXXX', 0.06), ('arpXXX', 0.04)] \\
feature	&	[('sahXXX', 0.26), ('yesXXX', 0.23), ('arpXXX', 0.04)]	&	[('sahXXX', 0.43), ('sunXXX', 0.29), ('yesXXX', 0.14)]	&	[('sahXXX', 0.26), ('yesXXX', 0.23), ('arpXXX', 0.04)] \\
in-mail	&	[('sahXXX', 0.12), ('arpXXX', 0.09), ('sunXXX', 0.09)]	&	[('sahXXX', 0.82), ('sunXXX', 0.18)]	&	[('sahXXX', 0.12), ('arpXXX', 0.09), ('sunXXX', 0.09)] \\
notification	&	[('karXXX', 0.42), ('sahXXX', 0.08), ('yesXXX', 0.06)]	&	[('karXXX', 0.82), ('sahXXX', 0.1), ('arpXXX', 0.04)]	&	[('karXXX', 0.42), ('sahXXX', 0.08), ('yesXXX', 0.06)] \\
oidc	&	[('sahXXX', 0.16), ('arpXXX', 0.11), ('yesXXX', 0.11)]	&	[('sahXXX', 0.65), ('yesXXX', 0.2), ('sunXXX', 0.1)]	&	[('sahXXX', 0.16), ('arpXXX', 0.11), ('yesXXX', 0.11)] \\
payment	&	[('arpXXX', 0.41), ('yesXXX', 0.41), ('surXXX', 0.18)]	&	[('surXXX', 0.47), ('sahXXX', 0.27), ('arpXXX', 0.13)]	&	[('arpXXX', 0.41), ('yesXXX', 0.41), ('surXXX', 0.18)] \\
reporting	&	[('surXXX', 0.32), ('sunXXX', 0.32), ('yesXXX', 0.31)]	&	[('surXXX', 0.5), ('sunXXX', 0.5)]	&	[('surXXX', 0.32), ('sunXXX', 0.32), ('yesXXX', 0.31)] \\
scheduler	&	[('surXXX', 0.11), ('sahXXX', 0.1), ('yesXXX', 0.02)]	&	[('surXXX', 0.5), ('sahXXX', 0.43), ('sunXXX', 0.03)]	&	[('surXXX', 0.11), ('sahXXX', 0.1), ('yesXXX', 0.02)] \\
search	&	[('surXXX', 0.29), ('sunXXX', 0.27), ('arpXXX', 0.26)]	&	[('sahXXX', 0.5), ('surXXX', 0.25), ('yesXXX', 0.12)]	&	[('surXXX', 0.29), ('sunXXX', 0.27), ('arpXXX', 0.26)] \\
survey	&	[('sahXXX', 0.28), ('arpXXX', 0.08), ('sunXXX', 0.07)]	&	[('sahXXX', 0.87), ('yesXXX', 0.09), ('arpXXX', 0.02)]	&	[('sahXXX', 0.28), ('arpXXX', 0.08), ('sunXXX', 0.07)] \\
task	&	[('surXXX', 0.19), ('sahXXX', 0.13), ('arpXXX', 0.1)]	&	[('surXXX', 0.67), ('sahXXX', 0.28), ('sunXXX', 0.06)]	&	[('surXXX', 0.19), ('sahXXX', 0.13), ('arpXXX', 0.1)] \\
user-tenant	&	[('sunXXX', 0.38), ('yesXXX', 0.25), ('arpXXX', 0.19)]	&	[('sunXXX', 0.54), ('surXXX', 0.17), ('sahXXX', 0.12)]	&	[('sunXXX', 0.38), ('yesXXX', 0.25), ('arpXXX', 0.19)] \\
video	&	[('karXXX', 0.13), ('sahXXX', 0.1), ('sunXXX', 0.06)]	&	[('sahXXX', 0.4), ('karXXX', 0.4), ('sunXXX', 0.15)]	&	[('karXXX', 0.13), ('sahXXX', 0.1), ('sunXXX', 0.06)] \\
    \hline
    Whole & [('surXXX', 0.1), ('yesXXX', 0.09), ('sunXXX', 0.09), & [('sahXXX', 0.36), ('yesXXX', 0.19), ('surXXX', 0.15), & [('surXXX', 0.1), ('yesXXX', 0.09), ('sunXXX', 0.09), \\ 
    
    Project &('sahXXX', 0.07), ('arpXXX', 0.07), ('souXXX', 0.07), & ('sunXXX', 0.13), ('aksXXX', 0.1), ('karXXX', 0.03), & ('sahXXX', 0.07), ('arpXXX', 0.07), ('souXXX', 0.07), \\
    
    &('841XXX', 0.05), ('aksXXX', 0.05), ('karXXX', 0.05)]& ('arpXXX', 0.03), ('ajaXXX', 0.01)] & ('841XXX', 0.05), ('aksXXX', 0.05), ('karXXX', 0.05)]\\
    
    \hline
    \end{tabular}
    }
    \caption{The Key Developers of Each Microservice and Whole Project (2024-11-21)}
    \label{tab:kdtable}
\end{table*}

\subsection{Context}

As context, we considered an open-source microservice-based project named \textit{ARC} by SourceFuse \cite{arc}. ARC is a rapid application development framework that supports the development of modern cloud-native applications at the enterprise level. It provides prebuilt microservices that can be deployed as a standardized reference architecture. Till 2024-11-21, the project has 16 independent microservices, the list of which can be seen in the ``services" folder of the repository \footnote{https://github.com/sourcefuse/loopback4-microservice-catalog/tree/master/services}. This project is commonly selected for studies on microservice architecture in terms of the developers' behaviors \cite{amoroso2023one,amoroso2024understanding}. Despite being a relatively young and small project, it provides sufficient data to provide quick and meaningful demonstration without imposing overloaded manual effort and risk in preprocessing (e.g., developer email mapping).

\subsection{Data Collection and Data Analysis}
We collected all the \textbf{1331 commits} from \textbf{61 contributors} and the \textbf{2194 issues} \textbf{from 2020-05-13 to 2024-11-21} using GitHub APIs 
For each commit, we further crawled the details regarding the contribution types and values in each file that the commit changes 
therein when 25282 commit change activities were collected. The collected commit information includes commit SHA, author ID (i.e., developer ID), created date, changed file, action type (i.e., add, modify, rename, or remove), and contribution amount 
(i.e., LOC). Regarding the issues, we also further crawled the details in each issue item, obtaining all the relevant activities in its timeline, e.g., commit SHA, author ID, etc. Only in the issue timeline can we trace it to the related commits with SHA toward constructing the artifact traceability graph. Furthermore, within all the developer IDs, we filtered out the bots from external tools, e.g., \textit{SNYK} (i.e., github+bot@snyk.io) and \textit{Dependabot} (i.e., dependabot[bot]@users.noreply.github.com). 
In addition, we found that many developers commit to the repository using multiple accounts. We manually unified 50 developers' alternative email accounts to obtain accurate results by checking their first name, last name, and initials combinations. This phenomenon commonly exists in most GitHub repositories. The data analysis uses the basic statistical approaches. 


\section{Results}
\label{sec:Results}
This section presents the results to answer the RQs.

\subsection{Key Developers (RQ$_1$)}

By adopting the identification methods described above, we identify the three types of key developers for each defined microservice by selecting only the committing and issue-handling activities for each specific microservice. To protect their privacy, 
we only show the initial three letters of their developer ID.  Regarding the methods, we adopted all the default parameters given by the replication package of \cite{ccetin2020identifying}. We also used their experiment's default sliding window of 365 days to demonstrate our results and to ease the comparison of annual changes.
Table \ref{tab:kdtable} shows each type's Top 3 local key developers for each microservice (considering the limited space) and the global key developers for the whole project. 
We can easily observe that the top global key developers (regardless of the developer types) are usually the local key developers for several microservices. For example, global \textit{Jack} \textit{surXXX} is also the top \textit{Jack} for five microservices. The local \textit{Jack} for the other microservices are also among the Top 5 global \textit{Jack}, e.g., \textit{yesXXX} and \textit{sunXXX}. All the Top 3 local \textit{Jack} of all microservices are global \textit{Jack} for the whole project as well. However, there are two developers, \textit{ajaXXX} and \textit{swaXXX}, who are global \textit{Jack}, but they are only the local \textit{Jack} for one microservice, respectively. 

We can also observe that the phenomena for \textit{Maven} and \textit{Connector} are the same, where global \textit{Maven} and \textit{Connector} are most likely covering many microservices as local\textit{Maven} and \textit{Connector}. Interestingly, we find that for 12 out of the 16 microservices, the Top local \textit{Jack}, \textit{Maven}, and \textit{Connector} are identical when it is the same for the global roles despite slight ranking differences. Importantly, we notice that for all microservices, the top \textit{Jack} and \textit{Connector} are the same developers even with the same ranks. However, there are some local \textit{Maven } not acting as top \textit{Jack} or \textit{Connector}, e.g., \textit{sahXXX} for the \textit{audit} and \textit{search} microservice. 

Compared with the key developer table for the same time in 2023 (in the replication package), we can observe that the situation was almost identical, with all microservices having the same local \textit{Jack} and \textit{Connector} and each developer covering multiple microservices as multiple roles (types). However, it is also worth noticing that many key developers in 2023 are not currently taking key roles anymore, e.g., \textit{shuXXX}, \textit{shiXXX}, and \textit{ragXXX}.

\subsection{Organizational Coupling (RQ$_2$)}

By adopting the evaluation method described above, we calculate the organizational coupling contributed by each developer from 2023-11-21 to 2024-11-21. For this period of time, five developers contributed to 1238.5 organizational coupling in total. Taking the standards from \cite{li2023evaluating}, the amount of coupling in one year is slightly over the limit of being ``heavily coupled''. However, in the total coupling of the whole project (115595.65 in total), the five developers perform effectively by creating only about 1.07\%, which shows the project is becoming more effectively organized. On the other hand, the five developers who cause organizational coupling are all key developers, while the other seven ``non-key'' developers did not cause any coupling. 

\begin{table}[!ht]
    \centering
    \footnotesize
    \begin{tabular}{l|r|r|r|r}
    \hline 
       \textbf{Dev.ID} & \textbf{1st Year} & \textbf{2nd Year}& \textbf{3rd Year}& \textbf{4th Year} \\
    \hline 
        samXXX & 11280.07& 35008.44.07 & 0& 0\\
        harXXX& 7001.41& 0& 0& 0\\
        ashXXX& 6652.26& 0& 0& 0\\
        aksXXX& 2659.22& 14043.83& 0& 0\\
        akaXXX& 2570.63& 0& 0& 0\\
        sumXXX& 1208.73 & 987.66 & 0& 0\\
        jamXXX& 161.37& 0& 0& 0\\
        yesXXX& 39.06& 0& 0& 0\\
        vaiXXX& 2.00& 0& 0& 0\\
        ankXXX& 0 & 6897.28 & 1785.52 & 0\\
        jyoXXX& 0 & 2725.30 & 0& 0\\
        barXXX& 0& 1932.01& 0& 0\\
        shuXXX& 0& 0 & 941.84 & 0\\
        ragXXX& 0& 0 & 252.88 & 0\\
        surXXX& 0& 11792.55 & 1212.00& 478.21\\
        sunXXX& 0& 0& 0& 359.68\\
        yesXXX& 0& 734.88 & 4317.93 & 228.47\\
        sahXXX& 0& 0& 0& 153.32\\
        arpXXX& 0& 0& 3330.48& 18.83\\
    \hline 
    \end{tabular}
    \caption{Key-developer-contributed OC Changes by Years}
    \label{tab:kdocchanges}
\end{table}

Furthermore, we also investigate the changes in the organizational coupling caused by the identified key developers through the project's lifecycle. Shown in Table \ref{tab:kdocchanges}, we can easily observe the changes in the key developers (only the ones causing OC) and the changes in the overall OC values. Each year, new developers start to take on key developer roles and, meanwhile, cause organizational coupling. However, similar to the observation from \cite{li2023evaluating} regarding the case of \textit{Spinnaker}\footnote{https://spinnaker.io/}, the overall OC values decreased sharply through the years, considering the numbers of commits are similar in each period. Meanwhile, the key developers contributed to 97.68\% of all the organizational coupling through the lifecycle, while only two non-key developers contributed the other 2.32\%. It should be noted that no clear correlation has been detected between organizational coupling and key developer metrics. 


\section{Discussion and Implications}
\label{sec:Discussion}
The obtained preliminary results show that, from any microservice-based project, we can identify the three types of key developers, i.e., \textit{Jack}, \textit{Maven}, and \textit{Connector}, in each microservice and across the whole project via the project's commit and issue data (RQ$_1$). With the selected case, we can observe the potential phenomenon that it is common for many key developers to take two or three roles (types) mentioned above. Meanwhile, the key developer team can change dramatically through the project's lifecycle as previous key developers leave the team regardless of the developer type. Meanwhile, we can also observe from the results that the identified key developers are the main contributors (97.68\%) to organizational coupling in microservice projects (RQ$_2$). However, it is also noticeable that organizational coupling has decreased over the years regardless of the commit numbers. 

Such early results provide an interesting perspective that the key developers are most likely the ones causing the organizational coupling, resulting in the potential degradation in microservice architecture. Even though this paper does not verify such an implication with one case, such a phenomenon must draw attention to the domain. Firstly, our next work is to apply the same approaches to a large number of microservice projects to verify the conclusion, e.g., using the 378 projects from \cite{amoroso2024dataset}. \textcolor{black} {With such a conclusion verified, future studies shall focus on the approaches and framework to optimize the microservice organizational structure towards decoupling teams and the monitoring and mitigation techniques preventing microservice degradation}. The potential research directions include but not limited to:


\begin{itemize}

    \item \textbf{Organizational Decoupling.} We need to propose an automated decoupling method to recommend and encourage (key) developers to focus more on the microservice(s) they are assigned and/or specialized. By doing so, the organization shall be gradually optimized, and the method will be more feasible and effective.

    \item \textbf{Degradation Monitoring.} Together with the organizational structure continuously optimized, the monitoring strategy for overall architecture degradation will be conducted by taking into account the static, dynamic, and organizational perspectives \cite{abdelfattah2024assessing,al2022using,li2024framework}.   

    \item \textbf{Developer Profile Compatibility.} To enhance the effectiveness of collaboration, the personality and tech roles, as part of the developers' profiles, should be considered and involved in the decision-making process for organization optimization \cite{li2024toward,amoroso2023one}.

    \item \textbf{Gamification as Motivational Affordance.} To conduct an effective decoupling strategy to motivate developers to focus on the tasks in particular microservices, gameful design and gamification can be used as motivational affordance \cite{li2024framework}. For such a purpose, the connection between the developer profiles and their player profiles can enable effective personalized design \cite{li2019statistical,li2024data}.

    \item \textbf{Generative AI and Tooling.} A comprehensive tool is to be implemented supporting such practices, which can also be adapted with special industrial interests considered. Especially with the advanced development of large language models and generative AI, the tool can be expected to offer enhanced performance and adaptability. 

\end{itemize}

The concrete framework will undeniably impact an industry where most software architectures face degradation issues, inevitably leading to software quality declination \cite{perry1992foundations}. The tool shall perform closely to the industrial best practice and adapt smoothly to their CI/CD pipeline. 


\textbf{Limitations.} We identified some limitations for this work. Firstly, the early result using only one case is limited in generalization, which will be validated in future studies with a larger number of cases of variety. Secondly, manual preprocessing of alternative developers' IDs can also impose a risk to legitimate results, especially for larger-scale projects in future studies.
 
\section{Related Works}
\label{sec:RW}

Many studies have examined the developer types and key developers based on different criteria. For example, Cheng and Guo~\cite{cheng2019activity} use factor analysis based on 19 different activity metrics to identify the active and supporting contributors. Di Bella et al.~\cite{di2013multivariate} also classify Core, Active, and Occasional developers using factor analysis. Cetin and T\"uz\"un~\cite{ccetin2020identifying,ccetin2022analyzing} provide a classification of the key developers considering their unique key contributions. However, no studies have contributed to the key developers in microservice-based projects toward optimizing coupling with such knowledge.

The coupling in (innovating) organizations was earlier interpreted as an achievement of coordination with a balanced degree of integration and specialization, where ``loose coupling'' is the optimal state \cite{lawrence1967differentiation}. Compared to the classic notion of ``high cohesion loose coupling'' in the software engineering domain, ``loose coupling'' has also been widely studied in the administrative science domain \cite{brusoni2001knowledge}. Regarding the coupling issue in microservice, Zhong et al. propose the microservice coupling index to measure how coupled microservices are at the static level \cite{zhong2023measuring}. Li et al.~\cite{li2023evaluating} focus on the coupling at the organizational level and measure the organizational coupling between microservices as the developers' frequent cross-service contribution. 



\section{Conclusion}
\label{sec:Conclusion}

This paper investigates the approach to identify key developers in microservice-based projects and their connection to the organizational coupling. \camera{The early result shows the feasibility of the approach and the potential for its generalizability.} It is a critical step toward the future concrete framework of microservice organizational structure optimization and the strategy of microservice degradation monitoring.

\section*{Acknowledgment}
This work was supported by the Academy of Finland (grant n. 349488 - ``MuFAno'') and Business Finland (``6G-Bridge 6GSoft''). This material is also based on work supported by the National Science Foundation (NSF) (Grant No. 2409933).

\bibliographystyle{IEEEtran}
\bibliography{Main}

\end{document}